# Direct observation of low frequency confined acoustic phonons in silver nanoparticles: Terahertz time domain spectroscopy

Sunil Kumar,[1,2] N. Kamaraju,[1,2] B. Karthikeyan,[1] M. Tondusson,[3] E. Freysz,[3] and A. K. Sood[1,2,*]

[1]*Department of Physics, Indian Institute of Science, Bangalore 560 012, India*

[2]*Center for Ultrafast Laser Applications, Indian Institute of Science, Bangalore 560 012, India*

[3]*Université de Bordeaux1, CPMOH, UMR CNRS 5798, 351, Cours de la liberation, 33405 Talence cedex, France*

Terahertz time domain spectroscopy has been used to study low frequency confined acoustic phonons of silver nanoparticles embedded in poly(vinyl alcohol) matrix in the spectral range of 0.1 to 2.5 THz. The real and imaginary parts of the dielectric function show two bands at 0.60 THz and 2.12 THz attributed to the spheroidal and toroidal modes of silver nanoparticles, thus demonstrating the usefulness of terahertz time domain spectroscopy as a complementary technique to Raman spectroscopy in characterizing the nanoparticles.

*Keywords:* Terahertz time domain spectroscopy, nanoparticles, dielectric function, low frequency phonons

## I. INTRODUCTION

Confined acoustic phonons in nanoparticles are fingerprints of their shape and size. The frequencies of these phonon modes are proportional to longitudinal/transverse sound velocity in the material and inversely proportional to the size of the nanoparticles. These modes have been studied extensively by low frequency Raman spectroscopy [1-7] and ultrafast laser pump-probe spectroscopy [8-12]. Lowest order spheroidal radial and quadrupolar modes [1-3,5,6,8] as well as their higher orders [4,5,11] have been observed for spherical nanoparticles. Though low frequency Raman scattering is used to study the modes with frequency down to 30 cm$^{-1}$, the peak positions and linewidths of very low frequency modes (below 30 cm$^{-1}$) are very difficult to measure accurately due to the high background from Rayleigh scattering. In a recent study [13], far infrared absorption spectroscopy was used to characterize titanium oxide nanoparticles by observing the spheroidal dipolar acoustic modes, the only study, to best of our knowledge, on experimental observation of infrared active vibrational modes in nanoparticles. Alternately, terahertz (THz) spectroscopy using picosecond and sub-picosecond terahertz pulses has found widespread applicability in recent years, particularly in spectroscopy of solid state materials [14-20] and biological systems [21-23], in the spectral range of 0.1-10 THz. Other than the nondestructive characterization of materials, THz time domain spectroscopy (THz-TDS) offers the contactless measurement of electrical conductivity [14]. The foremost advantage of THz spectroscopy is the simultaneous measurement of absorption and refraction and hence the complete optical and electrical characteristics without invoking the Kramers-Kronig analysis.

In this paper, we report on the observation of two confined acoustic phonon modes with frequency ~ 0.60 THz and ~ 2.12 THz in silver nanoparticles of average diameter ~ 3.7 nm embedded in poly(vinyl alcohol) (PVA) matrix by using THz-TDS. These observed frequencies are compared with the estimated frequencies of the spheroidal and the toroidal vibrational modes of a free elastic sphere in the Lamb's model [24-26]. The experimentally measured real and imaginary parts of the dielectric function of the nanoparticles-doped film show another resonance band centered at ~ 1.11 THz, which is attributed to the longitudinal acoustic mode (LAM) of vibrations associated with the straight segments of the long polymeric chains in the film [27].

## II. EXPERIMENTAL

The polymer embedded silver nanoparticle film (Ag-PVA) was cast from the mixture of aqueous solutions of PVA and high purity silver nitrate ($AgNO_3$) (both purchased from Aldrich chemicals). Solution of PVA was obtained by adding 3 gm of granular PVA in 30 ml of deionized water followed by continuous heating at 90 $^0$C and stirring for one hour. 5 mg of $AgNO_3$ was dissolved in 5 ml of deionized water to obtain aqueous solution of $AgNO_3$. The two solutions were then mixed and stirred for 20 minutes at 70 $^0$C. This mixture was poured into a petry dish and left undisturbed in a clean environment for a few weeks to dry up under ambient conditions. Thus obtained film was approximately 165 micron in thickness with silver content of ~ 0.17 wt%.

The Ag-PVA film contains Ag nanoparticles with nearly spherical shape as shown in Fig. 1(a) studied by transmission electron microscopy (TEM). The statistical size distribution of the particles is shown by the histogram in Fig. 1(b). The fitting of the TEM data with a Gaussian distribution gives the average diameter $d$ = 3.7±0.15 nm with variance $\sigma$ ~ 1 nm. In Figs. 1(c) and 1(d), the UV-Visible absorption spectra of the Ag-PVA film and a pristine PVA film of thickness ~ 380 micron obtained directly from the aqueous solution of PVA, are shown. The PVA film shows two small absorption bands at ~ 280 nm and ~ 335 nm. These features arise due to the presence of carbonyl containing structures connected to the PVA polymeric chains, mainly at the ends [28,29]. The intense broad band at ~ 420 nm for the Ag-PVA film is associated with the surface plasmon of the silver nanoparticles. The peak position of the surface plasmon absorption band of spherical metal nanoparticles is known to be only weakly size-dependent [30-32] and within the framework of free electron model, the size is inversely proportional to the spectral width of the surface plasmon peak [31].

In the THz-TDS we have used photoconductive emitter and detector (EKSPLA), both driven by 100 fs laser pulses with central wavelength at 800 nm from a 76 MHz Ti:sapphire laser. The set-up generates THz radiation in the spectral bandwidth of 0.1 to 2.5 THz with good signal to noise ratio of ~ $10^3$ at 0.8 THz.

## III. RESULTS AND DISCUSSION

The time domain electric fields, $E_s(t)$ and $E_r(t)$, associated with the THz pulse, recorded with and without the film are shown in Fig. 2(a). Fast Fourier transform gives direct access to the amplitude and phase at different spectral components of the THz pulse. Therefore, from THz-TDS, the frequency dependent absorption and refraction coefficients are obtained simultaneously without invoking Kramers-Kronig analysis. The spectral transmission coefficient, $T(\omega)$ of the film is the ratio between the complex spectral fields, $E_s(\omega)$ (with the film) and $E_r(\omega)$ (without the film) as

$$T_{\exp}(\omega) = \frac{E_s(\omega)}{E_r(\omega)} = A(\omega)\exp[i\phi(\omega)] \tag{1}$$

where $A(\omega)$ is the amplitude and $\phi(\omega)$ is the phase of the transmission coefficient. The spectral transmission coefficient of an optically thin film can be written as [33],

$$T_{th}(\omega) = \frac{\dfrac{4n^*(\omega)}{[n^*(\omega)+1]^2}\exp\{i\omega[n^*(\omega)-1]t/c\}}{1 - \dfrac{[n^*(\omega)-1]^2}{[n^*(\omega)+1]^2}\exp[i2\omega n^*(\omega)t/c]} \tag{2}$$

Here $n^*(\omega) = n(\omega) + iK(\omega)$; $n(\omega)$ is the index of refraction and $K(\omega)$ is extinction coefficient at angular frequency $\omega$, $t$ is the thickness of the film and $c$ is the speed of light in vacuum. Eq. 2 is compared with Eq. 1 and solved numerically for $n(\omega)$ and $K(\omega)$ which are finally used to calculate the real and imaginary parts of the complex dielectric function, $\varepsilon(\omega) = \text{Re}(\varepsilon) + i\text{Im}(\varepsilon)$ as $\text{Re}(\varepsilon) = n^2 - K^2$ and $\text{Im}(\varepsilon) = 2nK$. The experimental results for $\text{Re}(\varepsilon)$ and $\text{Im}(\varepsilon)$ of the Ag-PVA film are presented by filled circles in Figs. 2(b) and 2(c). These results have been fitted with a Lorentzian oscillator model given by,

$$\varepsilon(\omega) = \varepsilon_\infty + \sum_{k=1}^{3} \frac{F_k}{\omega_k^2 - \omega^2 - i\Gamma_k \omega} \tag{3}$$

Here $\omega_k/2\pi$ is the resonance frequency, $\Gamma_k/2\pi$ is the spectral width, $F_k$ is the oscillator strength, and $\varepsilon_\infty$ is the high frequency dielectric constant. The solid lines in Figs. 2(b) and 2(c) are the fits with fitting parameters given in Table I along with $\varepsilon_\infty = 2.75$. The resonance band centered at ~ 1.11 THz is similar to a band at ~ 1.25 THz observed for a pristine PVA film [27], as shown in the insets of Figs. 2(b) and 2(c). This mode is assigned to the longitudinal acoustic mode (LAM) of vibrations localized on the straight chain segments of the long polymeric molecules in PVA [27]. The other two resonance bands centered at ~ 0.6 THz and ~ 2.12 THz are attributed to the confined acoustic phonons in silver nanoparticles as discussed below. We note that the large spectral width of ~ 0.4 THz of these observed phonon modes could possibly be due to the large size distribution of the nanoparticles.

**A. Confined acoustic phonons in silver nanoparticles**

The vibrational motion of a homogeneous elastic sphere with free surface has been theoretically studied by Lamb [24], Nishiguchi *et al.* [25] and Tamura *et al.* [26]. Two sets of eigenfrequencies have been derived, the spheroidal (S) vibrational modes (with dilatation) and the toroidal (T) vibrational modes (without dilatation), given as

$$\omega_{l,m}^{S} = \xi_{l,m}^{S} \frac{2u_l}{d} \qquad (4a)$$

$$\omega_{l,m}^{S,T} = \eta_{l,m}^{S,T} \frac{2u_t}{d} \qquad (4b)$$

Here, $\xi_{l,m}$ and $\eta_{l,m}$ are the $(m+1)^{th}$ dimensionless eigenfrequencies with angular momentum quantum number $l$, $u_l$ and $u_t$ are the longitudinal and transverse sound velocities in bulk silver and $d$ is the diameter of the nanoparticle. Raman and infrared activity of these modes is being debated in the literature [34-38] and a quantitative measure of the scattering and absorption cross-sections of various modes is lacking. For the spheroidal modes, there is a general consensus that modes with even $l$ (0, 2, 4....) are Raman active and modes with odd $l$ (1, 3, 5…) are infrared active [1,34,35]. On the other hand, for the toroidal modes, Duval *et al.* [34,35] argue that there are no Raman active modes (both for even and odd $l$), whereas Fujii *et al.* [1] predict that modes with odd $l$ can be Raman active.

The lowest order ($m = 0$) and a few higher order ($m \geq 1$) Raman active, spheroidal radial ($l = 0$) and quadrupolar ($l = 2$) modes have been routinely observed either by Raman spectroscopy [1-7] or by ultrafast laser pump-probe spectroscopy [8-12]. On the other hand, only very few studies are available on experimental observation of the infrared active modes. The only observation we know of, is that of Murray *et al.* [13] who have observed the spheroidal dipolar acoustic modes ($l = 1$, $m \geq 0$) in titanium oxide nanoparticles by far infrared absorption spectroscopy: a single band at frequency 40 cm$^{-1}$ (1.2 THz) for nanoparticles with diameter $d = 4.5$ nm was attributed to the $m = 0$ mode, whereas single band at 48 cm$^{-1}$ (1.44 THz) for nanoparticles with diameter $d = 7.5$ nm was attributed to the $m = 1$ mode. It is important to note that the $l^{th}$ vibrational mode is ($2l+1$)-fold degenerate for a spherical nanoparticle which gets lifted for an elliptical particle [4,35,39] and hence can contribute to the spectral width of the observed modes. Some times, it is possible that a higher order mode is observable experimentally whereas the lower order is absent. For example, the surface modes ($m = 0$) are suppressed for particles embedded in a host matrix [4,26] and the spheroidal modes which do not change the aspect ratio are expected to be less coupled to the radiation [4,39].

For silver nanoparticles, using $u_l = 3650$ m/s and $u_t = 1660$ m/s [1], the calculated eigenfrequencies for various $m$ are given in Tab. II for spheroidal modes with $l = 1$ and toroidal modes with $l = 2$. The values of the sound velocities differ by about ~ 5 % in Ref. [40]. Such a difference will only make a similar change in the estimated frequencies of the modes. It can be seen that the observed band at 0.60 THz is close to the spheroidal mode with $m = 1$, band at ~ 1.11 THz is close to the spheroidal mode with $m = 2$ as well as toroidal mode with $m = 1$, and the band at ~ 2.12 THz is close to spheroidal mode with m = 5 as well as toroidal mode with $m = 3$. We may note that the continuum assumption in Lamb's theory for a homogeneous elastic particle may not hold exact for smaller particles [3]. Second, in Lamb's theory, the frequencies have been derived for a free sphere with stress free boundary conditions. Tamura *et al.* [26] observed that the mode frequencies are shifted upward when the host matrix effect is taken into account. More recently, calculations by Murray *et al.* [40] using various models for silver nanoparticles embedded in a glass matrix also showed that the mode frequencies blue-shift up-to ~ 10% as compared to those obtained using a free sphere model. Our present experimental observation of vibrational modes of silver nanoparticles at frequencies of ~ 0.6 THz and ~ 2.12 THz have a broad spectral width of ~ 0.4 THz due to the broad particle size distribution and hence free sphere model can suffice. Theoretical investigation of the Raman and infrared activity along with the comparative study of absorption and scattering cross-section of various possible modes is needed for quantitative understanding of the observed modes.

**IV. CONCLUSIONS**

In summary, we have studied the terahertz frequency response of silver nanoparticles embedded in a PVA matrix. For the first time, we have observed the confined acoustic phonons in silver nanoparticles by measuring the real and imaginary parts of the dielectric function of the nanoparticle-doped polymer film using THz-TDS. The observed phonons with frequency 0.6 and 2.12 THz are attributed to the spheroidal and toroidal vibrational modes. The LAM associated with the crystalline lamellae in PVA has been observed at frequency 1.11 THz. Our experiments demonstrate the use of THz-TDS as complementary technique to Raman scattering to characterize the nanoparticles. We note that the phonons observed in the THz-TDS are the equilibrium phonons of the system. The next step will be to study the time evolution of these modes of nanoparticles once created coherently in the ISRS by using an optical-pump. The THz absorption spectrum of the nanoparticles is measured with and without the pump as a function of time-delay between the optical-pump pulse and the THz-probe pulse in an optical-pump THz-probe experiment. It is possible that in the THz spectrum new modes can appear due to the pump excitation induced infrared activity of certain modes.

## ACKNOWLEDGEMENTS

AKS thanks Department of Science and Technology and EF thanks CNRS for financial assistance. SK acknowledges UGC for Senior Research Fellowship. This work was partially supported by Indo-French P2R program.

*Corresponding author: Electronic mail: *asood@physics.iisc.ernet.in*

TABLE I. Parameters obtained from fitting Eq. 3 with the experimentally obtained values of dielectric function for the Ag-PVA film of thickness 165 micron.

| mode $k$ | $\omega_k/2\pi$ (THz) | $\Gamma_k/2\pi$ (THz) | $F_k$ (THz$^2$) |
|---|---|---|---|
| 1 | 0.60±0.04 | 0.29±0.11 | 2.4±0.8 |
| 2 | 1.11±0.02 | 0.61±0.10 | 17.7±2.8 |
| 3 | 2.12±0.01 | 0.49±0.05 | 34.3±2.8 |

TABLE II. Frequencies in THz as a function of the mode index, $m$ of the IR active spheroidal (S), $l = 1$ and toroidal (T), $l = 2$ acoustic modes for a silver nanoparticle with diameter 3.7 nm. The calculations are followed from Refs. [24-26].

| $m$ | S ($l=1$) (THz) | T ($l=2$) (THz) |
|---|---|---|
| 1 | 0.50 | 1.00 |
| 2 | 1.02 | 1.47 |
| 3 | 1.36 | 1.93 |
| 4 | 1.52 | 2.40 |
| 5 | 1.96 | 2.87 |

**Figure captions**

FIG. 1. Characterization of the silver nanoparticles embedded in PVA matrix. (a) Transmission electron microscope image of the nanoparticles, and (b) their statistical size distribution. (c) UV-Visible absorption spectra of the Ag-PVA, and (d) the PVA films.

FIG. 2. (a) Time domain THz pulse with and without the Ag-PVA film, (b) Im($\varepsilon$), and (c) Re($\varepsilon$). Continuous lines in (b) and (c) are the fits using Eq. 3 with fitting parameters given in Tab. I. The locations of the confined acoustic phonons of silver nanoparticles at 0.6 THz and 2.12 THz are marked by vertical arrows (↑). The broad peak at 1.11 THz corresponds to the LAM of crystalline lamellae in the PVA film. The insets of (b) for Im($\varepsilon$) and (c) for Re($\varepsilon$) show the results for a pristine PVA film (Ref. 27).

Figure 1.

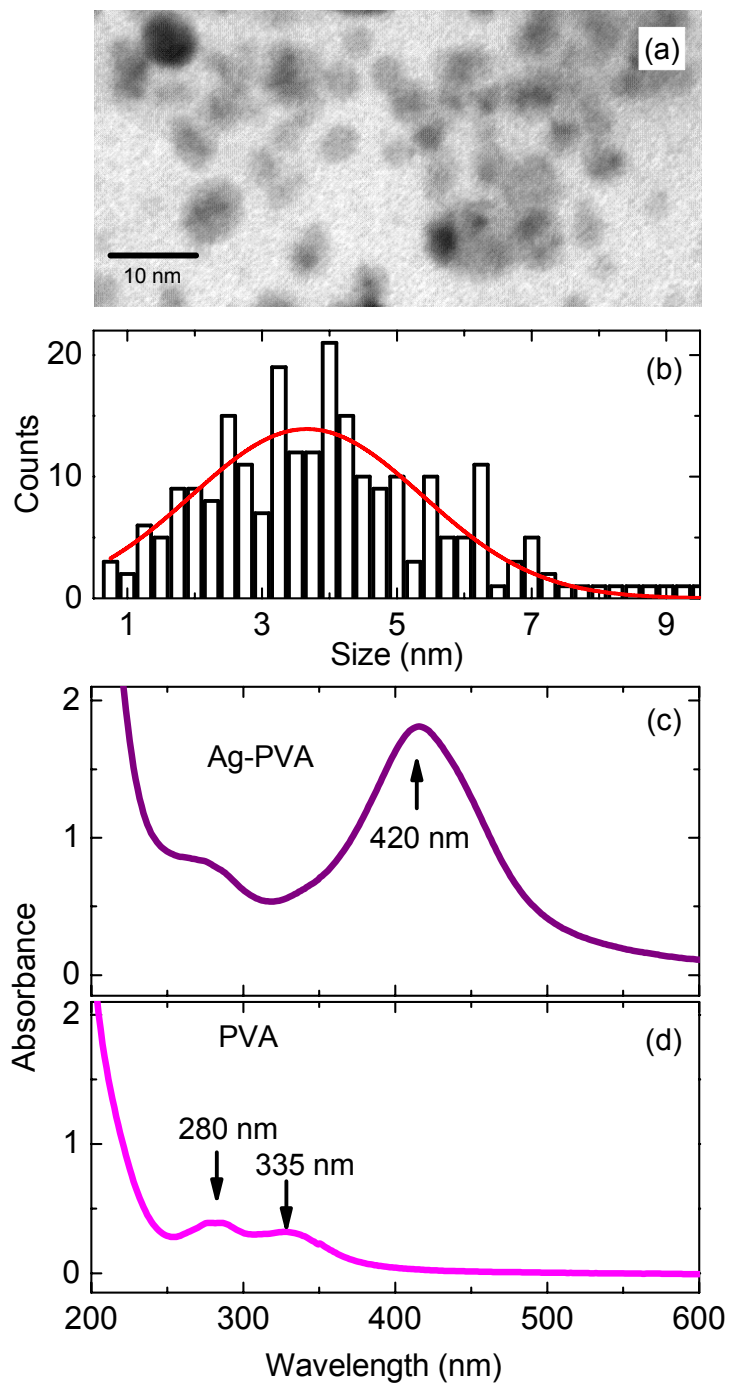

Sunil Kumar et al.

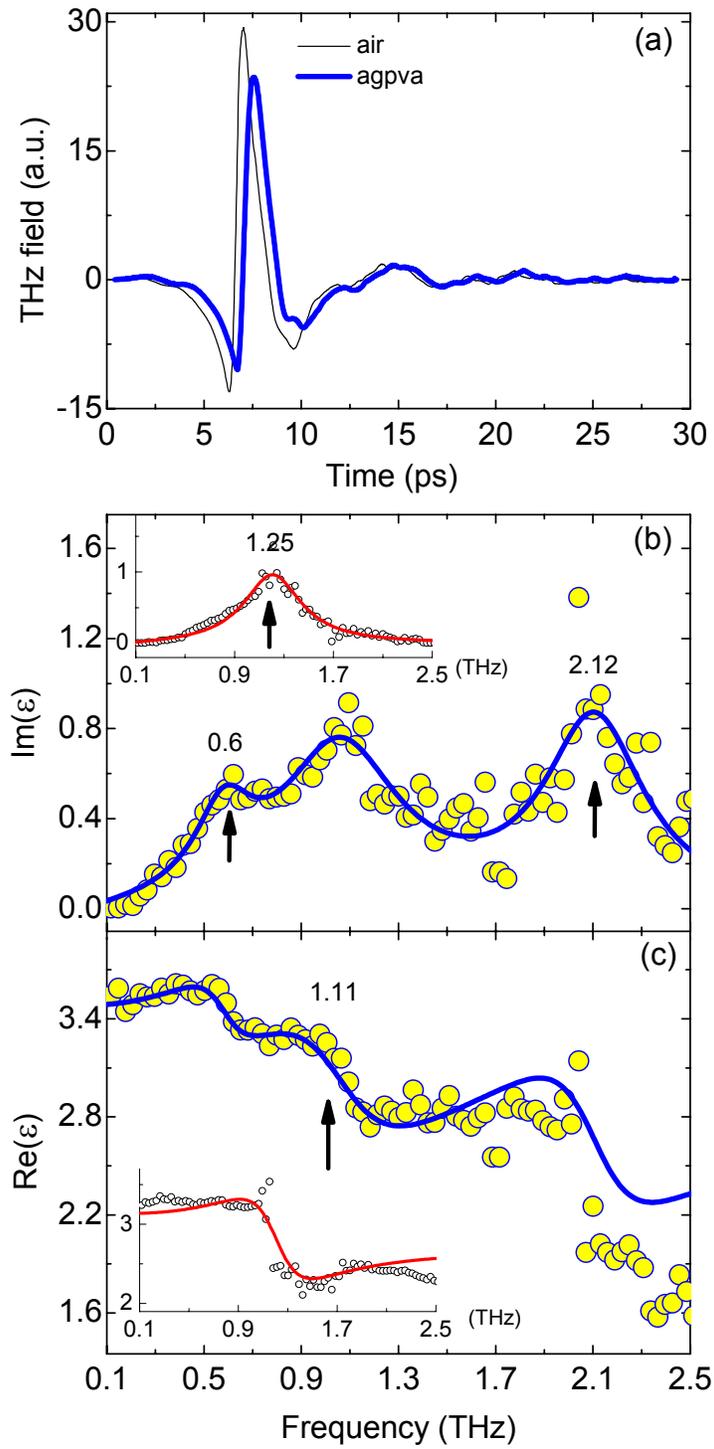

Sunil Kumar et al.